\documentclass[]{article}  
\usepackage{epsfig}              

\def\ompipi  {  \omega \pi^- \pi^0  }
\def\omrho   {  \omega \rho^-  }
\def\reactf  { $\pi^- Be \rightarrow \pi^+ 2 \pi^- 2 \pi^0 Be$ }
\def\reactl  { $\pi^- Be \rightarrow \omega \pi^- \pi^0 Be$ }
\def\at      {  a_2 (1320)  }
\def\aq      {  a_4 (2040)  }
\def\gev     { GeV }
\def\mev     { MeV }
\def\tprime  { t^{\prime} }
\def\tripi   { \pi^+\pi^-\pi^0 }
\def\fivpi   { \pi^+ 2 \pi^- 2 \pi^0 }

\def\jpme    { J^PM^{\eta}}

\begin{document}

\title{ \bf Study of the reaction \bf{\reactl}}   
\author{VES Collaboration \footnotemark}    
\date{14.01.98}    
\maketitle

\begin{abstract}
    The results of partial wave analysis of the $\ompipi$-system, 
produced in the reaction \reactf at beam momentum of $37 \, \gev$
are presented. 
An indication of the resonant type structure with mass 
$M \approx 1.74 \, \gev$ and $\jpme =0^-0^+$ in the $\omrho$-system was found.  
The decay branching ratio of $ \at^- $ to $\ompipi$ was measured.
A wave with $\jpme =2^+1^+$ shows a broad bump at mass
$ M \approx 1.7 \, \gev $. 
The decays $\pi_2(1670)^- \rightarrow \omrho$ and $ \aq^- \rightarrow 
\omrho $ were observed for the first time. 
The bump in the $b_1(1235)\pi$ wave with exotic quantum numbers $J^{PC}=1^{-+}$
at $ M \approx 1.7 \, \gev $ was observed.
\end{abstract}

\footnotetext[1]{D.V.Amelin, E.B.Berdnikov, G.V.Borisov,
V.A.Dorofeev, R.I.Dzhelyadin, A.V.Ekimov, Yu.P.Gouz,
I.A.Kachaev, A.N.Karyukhin, Yu.A.Khokhov, A.K.Konoplyannikov, S.V.Kopikov, 
V.V.Kostyukhin, S.A.Likhoded, V.D.Matveev, A.P.Ostankov,
D.I.Ryabchikov, A.A.Solodkov, O.V.Solovianov, E.A.Starchenko,
N.K.Vishnevsky, E.V.Vlasov, A.M.Zaitsev\\
Institute for High Energy Physics, 142284, Protvino, Russia\\
G.Sekhniaidze, E.G.Tskhadadze\\
Institute of Physics, 380077, Tbilisi, Georgia}

\newpage
\section{Introduction}             
	The results of partial wave analysis (PWA) 
of the $\ompipi$-system, produced in the reaction \reactl by pion beam with
momentum $P_{\pi}=37 \, \gev$ are presented in this article.
These results were partially reported at the conferences \cite{Manchester},
 \cite{Warsaw}.

	The study of this reaction is an interesting subject for several reasons.
The quark model predicts many meson resonances with masses in the region 
of $1.5-2.5 \, \gev$. The exotic mesons with the same masses may exist apart from
the ordinary $q \bar q$ states. The hybrids, four-quark mesons, hadron molecules
can be in the isovector state. Practically all of them should decay to
$\ompipi$, where the $\omrho, b_1\pi, \rho_3(1670)\pi$ should be the main chanels.
There exist in several cases definite predictions
for probabilities of the decay of different objects into these chanels, namely
$\omrho$. Study of these decays can play important role in 
determination of the nature of states under study, e.g. the results of
the model calculations for decay widths of the $\pi(1800)$ 
\cite{Kachaev},\cite{Kostyukhin}. That state can be a hybrid or a 
radial excitation of the $\pi$-meson, $3^1S_0$. The difference in the 
decay widths into $\omrho$ \cite{Close} predicted by the $^3P_0$ model makes
it possible to distinguish between these alternatives. Similar predictions
were made \cite{Close} for the $a_1(1700)$ state \cite{Kachaev}, which can 
be a hybrid or a $2^3P_1$ radial excitation state. 
The predictions for 
the $\pi_2(1670)$ and for more heavy states with $\jpme =2^-0^+$
\cite{Close} are worth mentioning.
 
	The system composed of two vector mesons has been previously
studied in the $\gamma-\gamma$ interactions  \cite{Gamma-gamma},
in the $J/\psi$ radiative decays\cite{Jpsi}, 
in the nucleon-antinucleon annihilation \cite{NNbar},
in the charge exchange processes \cite{exchange}, 
in the central \cite{centr.prod.} and diffractive production\cite{Manchester}.
The relatively small probability of the production processes and/or the high
multiplicity of the final states did not let them to take 
the experimental data sample sufficient for making detailed PWA.
Nevertheless, the indications to the existence of isoscalar resonances with 
quantum numbers $2^{++}$ in the $\rho\rho$ and $\omega\omega$ systems
were obtained.
	
	The experimental information about isovector mesons is even 
scarcer. The diffractive productions of the isovector resonances 
with $J^P=0^-,1^+,2^-...$ is especially suitable, for the production cross 
sections are large and the physical backgrounds are small. The waves with 
quantum numbers $J^P=2^+,4^+$ become important at relatively large momentum 
transfer.

\section{Set-up description}           

	The measurements were carried out using VES spectrometer exposed 
to the negatively charged pions with momentum $37 \, \gev$. 
VES is a wide aperture spectrometer of charged particles and photons. 
The beam part is composed of the threshold Cerenkov counters and a system 
of propotional chambers.
The vertex part consists of a 4cm thick Be target and a guard system 
surrounding it to veto events with particles moving backward to the beam 
in the event centre of mass system. 
The spectrometer of charged particles consists of a tracking 
system, a wide aperture magnet( kick momentum $0.7 \, \gev$ ) and a multichanel
threshold Cerenkov counter. A fine grained electromagnetic calorimeter
(1600 cells) measures $\gamma$ energy and position. A triggering system, which
consists of a hodoscope, scincillation counters and a guard system selects events of beam interaction with the target and at least two tracks in the forward
spectrometer. The data sample used for presented analysis included 
$  \sim 6 \cdot 10^7$  reconstructed events, corresponding to the integrated
luminosity of $ \sim 100 \, \mu b^{-1}$. The acceptance for events with several
charged and neutral pions is rather high ( 
$\epsilon \approx 0.6 \mbox{ for }\omega(\tripi)\pi^-\pi^0
\mbox{ at } M_{\omega \pi^- \pi^0}=1.5 \, \gev $ )  and
is a smooth function of kinematical variables. 
The measured width of the $\pi^0$ 
meson in the $\gamma\gamma$ invariant mass spectrum charactirises the 
experimental resolution of calorimeter and is equal to $ \sim 20 \, \mev$.
The measured width of the $\omega$-meson $\Gamma \sim 30 \, \mev$ includes 
the resolution of the tracking part of set-up.
    
\section{Event selection and main features of the reaction \reactl} 
	
	At first, the events of the process \reactf were selected for further
analysis of the $\ompipi$ system using the following criteria for
reconstructed events:
\begin{itemize}
\item Two negative and a positive track originated in the target region.
\item At least four energy clusters in the calorimeter
not associated with tracks. A cluster is assumed to be a $\gamma$-shower, if it
satisfied the criterion on the transverse form of the shower. Events with
the pairs of photons with invariant mass nearest to the PDG value
and within $ 60 \, \mev $ range for $\pi^0$ and $ 156 \, \mev$ for $\eta$
were selected.
\item Some $ \sim 10 \% $ of events with electrons were rejected using
the ratio of the energy deposited in calorimeter and the momentum measured by magnetic spectrometer.
\item The Cerenkov counter responce to passed track was applied for
charged $K$-meson identification. Some $3.4 \%$ of events were rejected
as having kaons.
\item Events with two $\pi^0$ and, possibly free soft photons $( E_{\gamma}
\leq 1 \, \gev)$, and the total forward final state particles energy in the range
$ 34 \, \gev \leq \sum E_i \leq 39 \, \gev $ were classified as those of type
\reactf.
\end{itemize} 
	
\begin{figure}
\epsfig{file=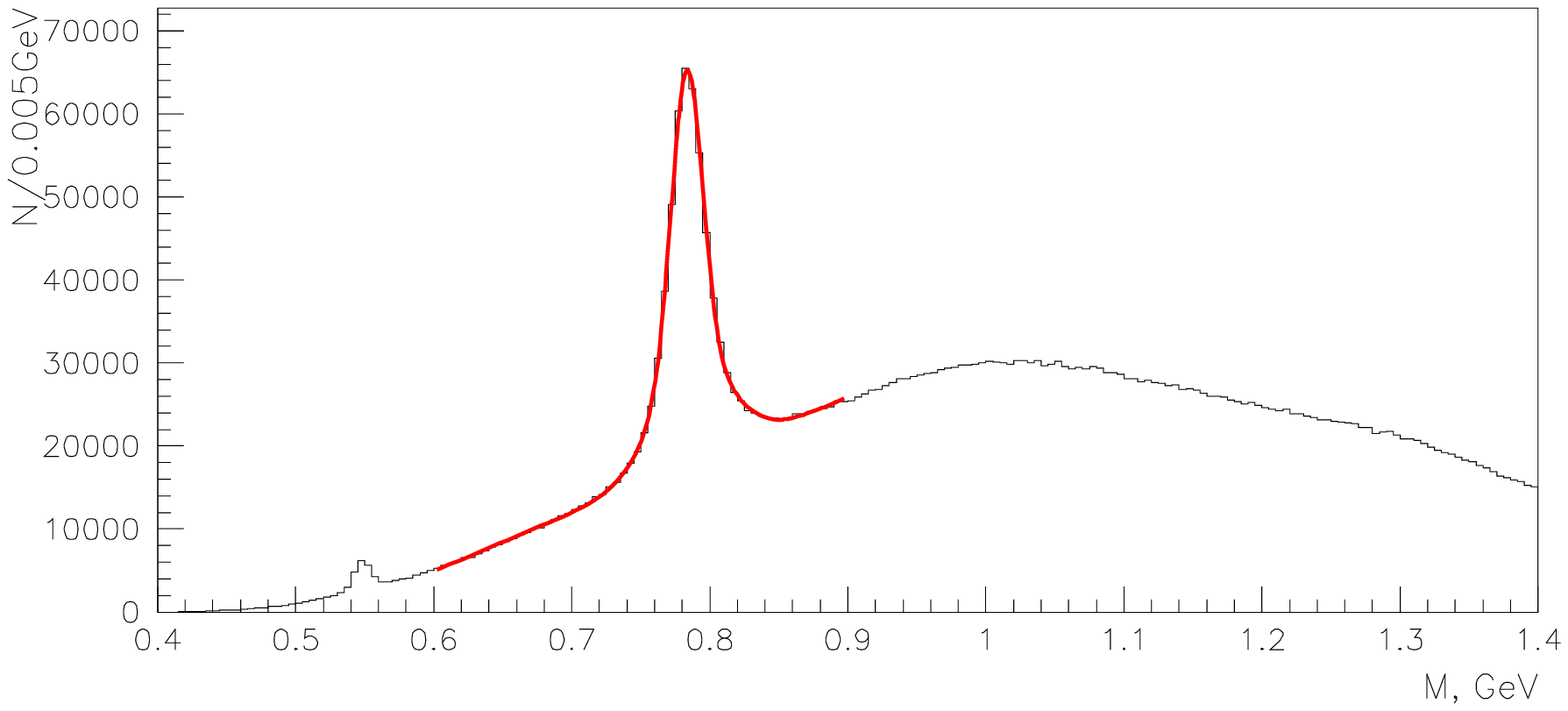,height=5cm,width=12cm,bbllx=0pt,bblly=288pt,bburx=530pt,bbury=530pt}
\caption{The invariant mass of the $\tripi$ for events of $\fivpi$ production.}
\label{omega3pi}
\end{figure}

	A 1C-fit was done on the mass of $\pi^0$. The invariant mass 
distribution of the $\tripi$ system with additionaly produced $\pi^-\pi^0$ 
is presented in fig.~\ref{omega3pi}. Clear peaks corresponding to 
the $\eta$ and $\omega$-meson production with large backgound are evident.
The $\omega$-meson peak has a non-gaussian form. Its parameters were
determined by the fit with the sum of two gaussian functions in the range from
$0.6 \, \gev $ to $ 0.9 \, \gev$. The mass of the narrow part was
$M_{\omega}= 783.37 \pm 0.06 \, \mev  $, where the systematic error was estimated to be  $\sigma_{M_{\omega}} = 1 \, \mev $. 
The width was:  $\Gamma _{\omega}= 26.2  \pm 0.2 \, \mev$ and the number of 
events with $\omega$ production turned out to be $(361 \pm  1)\cdot10^3$.
The fraction of background events at the point of the peak maximum was 
estimated to be $ \sim 30 \% $.

	To make various distributions describing the production and decay of 
the $\ompipi$ system the background was estimated by the neighbouring to 
the $\omega$ peak bands and was subtracted. 
Application of this method assumes smooth behaiviour of the background. 
The drawback is the difference of the phase spaces for those bands,
where the background was estimated, and for the signal ones. 

	Another method of background subtraction was applied for cross 
checking. The matrix element of the $\omega \rightarrow \tripi$ decay is 
propotional to:   
\begin{equation}
    \lambda = \frac{4 \cdot [ \vec{p}_{\pi^+} \times  \vec{p}_{\pi^-}]^2}{(M_{3\pi}^2- 9m_{\pi}^2)^2}
\end{equation}
where:

	$\vec{p}_{\pi^+}$ - momentum vector of $\pi^+$ in the $\tripi$ centre of mass,

	$\vec{p}_{\pi^-}$ - momentum vector of $\pi^-$ in the $\tripi$ centre of mass, 

	$M_{3\pi}$ - $\tripi$ invariant mass,

	$m_{\pi}$ - $\pi$ meson mass.

If the distribution of $\lambda$ for background events is sufficiently uniform,
then by weigthing events having $\lambda < 0.5 $ with -1 and those having 
$\lambda \geq 0.5 $ with +1 we effectively subtract background. 
But the efficient number of signal events filtered this way is twice smaller 
than that in the initial sample. 

\begin{figure}
\epsfig{file=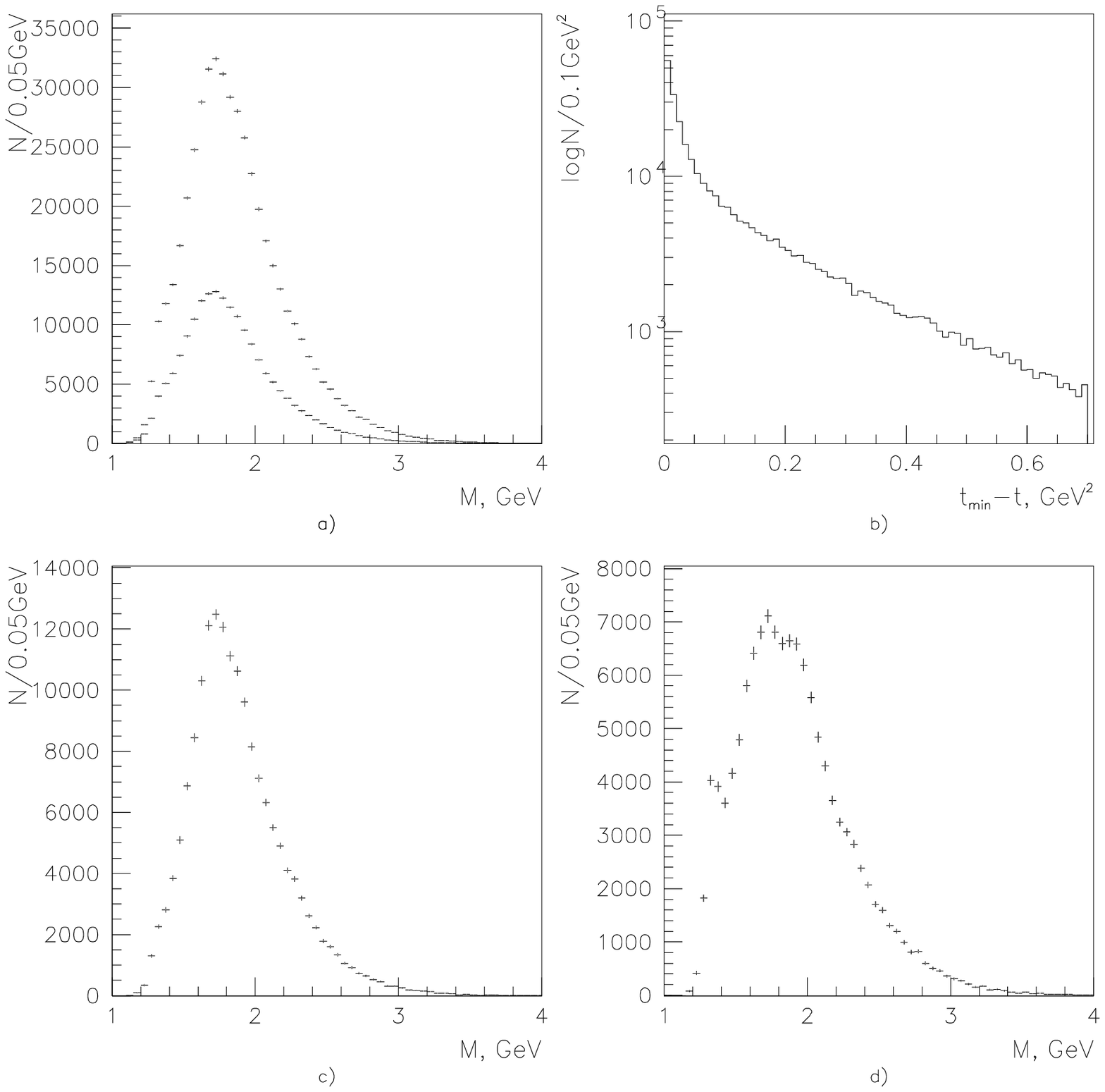,height=12cm,width=12cm,bbllx=0pt,bblly=0pt,bburx=530pt,bbury=530pt}
\caption{ a) Mass of the $\fivpi$-system with  
$-\tprime < 0.7 \, \gev^2 $ from the band $0.7577 \le M_{\tripi} < 0.8077
\, \gev$ with superimposed distribution from the control bands.
b) Momentum transfer $\tprime$ for reaction \reactl. Mass of the $\ompipi$
system for:  LT - c) and  HT - d).}
\label{om2pi_m_t}
\end{figure}

\begin{figure}[htpb]
\epsfig{file=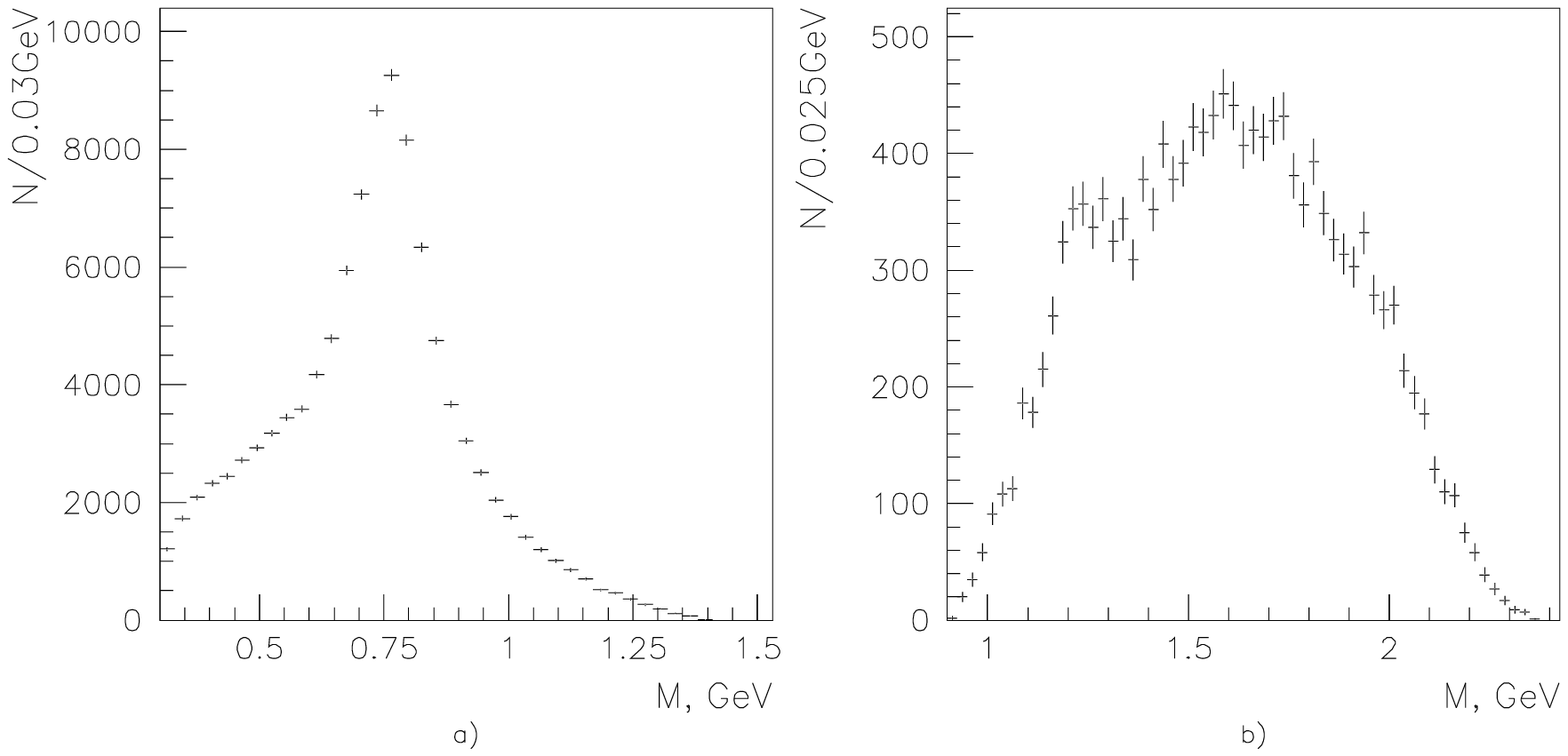,height=5cm,width=12cm,bbllx=0pt,bblly=288pt,bburx=530pt,bbury=530pt}
\caption{a) Mass of $\pi^-\pi^0$ with
$-\tprime < 0.08 \, \gev^2 $, $1.65 \, \gev \leq M(\fivpi) < 2.15 \, \gev$.
b)  Mass of $\pi^+2\pi^-\pi^0$ with
$-\tprime < 0.08 \, \gev^2 $, $2.15 \, \gev \leq M(\fivpi) < 2.45 \, \gev$.
}
\label{pair_mass}
\end{figure}

	Both methods resulted in statistically compatible distributions of
variables characterizing the $\ompipi$ decay and production. The results of the
subtraction procedure by the former method is presented in 
fig.~\ref{om2pi_m_t}. Two slopes are clearly seen in the momentum transfer 
squared distribution ($\tprime = t - t_{min}$) in fig.~\ref{om2pi_m_t}b.
It is useful to divide the experimental event sample into two parts in the variable
$\tprime$: type LT ( low $\tprime$ ) $-\tprime < 0.08 \, \gev^2 $ and 
HT ( high $\tprime$ ) $0.08 \leq -\tprime < 0.7 \, \gev^2$. 
The $\rho$-meson peak dominates in the $\pi^-\pi^0$ invariant mass spectrum
(fig.~\ref{pair_mass}a). The $b_1(1235)$ isobar of the $\omega\pi$-system is more 
evident in the high mass region of $\fivpi$ (fig.~\ref{pair_mass}b). 
The $\ompipi$ invariant mass spectrum of the LT type (fig.~\ref{om2pi_m_t}c)
has its maximum at $1.72 \, \gev$ and a weakly pronounced
shoulder in the region around $2 \, \gev$. The distribution of HT type events   
(fig.~\ref{om2pi_m_t}d) is wider and has a shoulder in the region of 
$2 \, \gev$.
A peak cooresponding to the $a_2(1320)$ meson production is clearly seen in the
low mass region.
	The PWA method was applied to study this complicated structure.
Further the event sample, where the mass of at least one of the $\tripi$
combination satisfies the relation
$|M_{\tripi}-0.7827|<0.060 \, \gev$, were analysed.

\section{Procedure of the partial wave analysis} 

	The goal of this method is the determination of intensities and 
relative phases of different states as a function of $M_{\ompipi}$ 
and $\tprime$. This method is based on the approach realised in 
Illinois PWA program  \cite{Hansen}.

	The full amplitude of each decay channel is written in the form of the 
product of production and decay amplitudes. 
The produced system decay process is assumed to
be a chain of sequential two body decays into the isobar and a bachelor meson. 

The $\ompipi$-system state is described at fixed $M_{\ompipi}$ and $\tprime$
by the following set of quantum numbers: the total angular momentum $J$ of 
$\ompipi$ system, the parity $P$, the projection of total angular momentum 
onto the 
Gottffried-Jackson reference frame $z$-axis $M$, the isobar orbital momentum in
the $\ompipi$ decay $L$ and the total spin $S$. It is convinient to change
the basis of states having $M = \pm 1$ to the basis of their linear 
combinations, which are the eigenvalues of a reflection operation in the 
production plane. The quantum number which characterises the
behaviour of the system 
with respect to reflecton is called a naturality, $\eta = \pm 1$. As a result
any partial wave is described by the following set of numbers: 
$\jpme L[S](isobar)$ with $M \geq 0$.

	A decay amplitude into the state with a definite orbital angular 
momentum $L$ and a total spin $S$ was written in the frame of a non-relativistic
tensor Zemach formalism \cite{Zemach}. The isobar mass dependence of the
decay amplitude was assumed in the relativistic Breit-Wigner form 
with the Blatt-Weisskopf form-factors \cite{BlattW}.

	Integrals of the amplitude products were calculated by 
the Monte-Carlo method using the set-up model.

	The selected events were divided into two parts in $\tprime$: the LT and HT, each of which, in turn, was divided into the $50 \, \mev$ and $100 \, \mev$ bins in $M_{5\pi}$, beginning from $M_{5\pi}=1.2 \, \gev$. 
	The process of the beam diffraction dominates in the region
of small momentum transfer, and this fact dictates the selected wave set with
$M^{\eta}=0^+$ and $J^P=0^-,1^+,2^-$ ,... The waves with $M^{\eta}=1^+$ 
and $J^P=2^+,4^+$ become dominant in the region of high momentum transfer. This
is proved by the $a_2(1320)$ peak in the $M_{5\pi}$ spectrum for events from the HT sample. 
These facts determined the wave set used in the PWA. The following
isobars were included: the $\rho^-$ in the $\pi^-\pi^0$ chanel, the $b_1 \mbox{, and } \rho_1
\mbox{, and } \rho_3(1690)$ in the $\omega\pi$ chanel. 
The $1^-1^+S(b^0_1)$ wave with 
indefinite isospin was included for the description of the visible $b_1^0$ prevailing
over $b_1^-$. The incoherent wave FLAT, uniformly distributed over the $5\pi$
phase space with one of the $\tripi$ combination from the $\omega$ region, was 
included to take the non $\ompipi$ background into account.

	The resultant wave set looked as following:
 
\begin{tabular}{lllllllllll}
FLAT &&&&&&&&&&\\
${\bf 0^-}$ && $0^-0^+\ P1(\rho)$ &&                      &&                    &&&\\
${\bf 1^+}$ && $1^+0^+\ S1(\rho)$ &&                      && $1^+0^+\ P1(b_1)$   &&\\  
${\bf 1^+}$ && $1^+0^+\ D1(\rho)$ && $1^+0^+\ D2(\rho)$   &&                     &&
 $1^+0^+\ S1(\rho_1)$   &\\  
${\bf 2^-}$ && $2^-0^+\ P1(\rho)$ && $2^-0^+\ P2(\rho)$   && $2^-0^+\ P1(b_1)$   &&
 $2^-0^+\ P1(\rho_1)$   &\\  
${\bf 3^+}$ && $3^+0^+\ D1(\rho)$ && $3^+0^+\ D2(\rho)$   && $3^+0^+\ P1(b_1)$   &&
 $3^+0^+\ D1(\rho_1)$   &\\
${\bf 3^+}$ && $3^+0^+\ S3(\rho_3)$ &&                    &&                    &&&\\  
${\bf 1^+}$ && $1^+1^+\ S1(\rho)$ && $1^+1^+\ D1(\rho)$   &&                    &&&\\  
${\bf 2^+}$ && $2^+1^+\ S2(\rho)$ &&                      &&                    &&&\\  
${\bf 3^+}$ && $3^+1^+\ D1(\rho)$ && $3^+1^+\ D2(\rho)$   &&                    &&&\\  
${\bf 4^+}$ && $4^+1^+\ D2(\rho)$ &&                      &&                    &&&\\  
${\bf 1^-}$ &&                    &&                      && $1^-1^+\ S1(b_1)$  &&&\\  
${\bf 1^-}$ &&                    &&                      && $1^-1^+\ S1(b_1^0)$ &&&\\  
\end{tabular}
\newline

\begin{figure}[htpb]
\epsfig{file=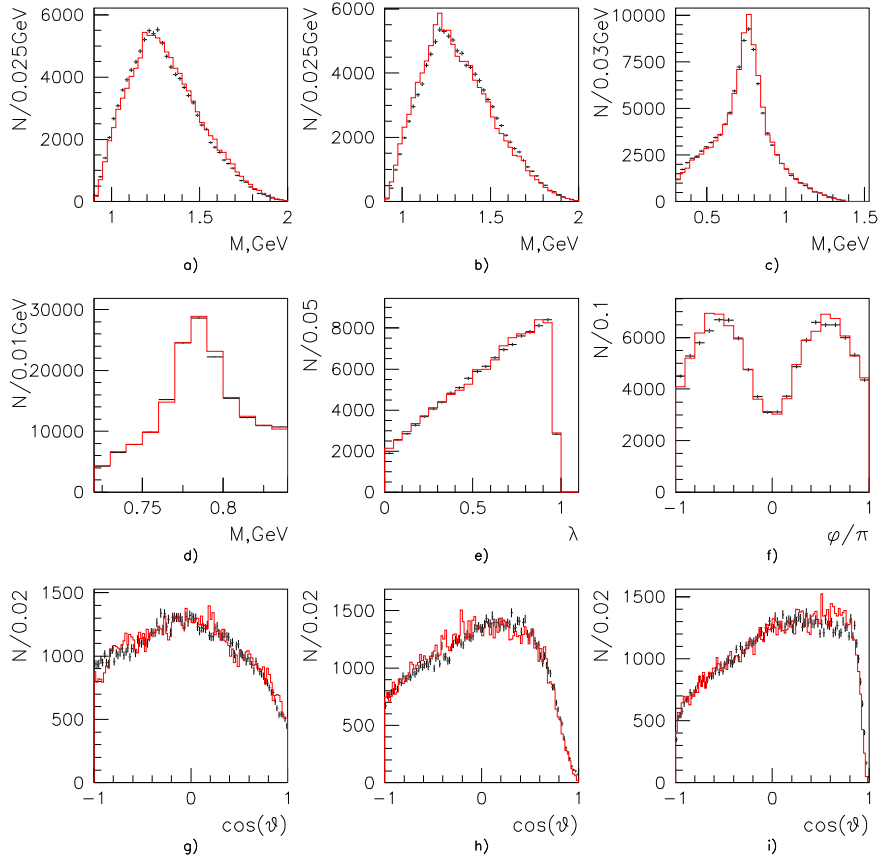,height=12cm,width=12cm,bbllx=0pt,bblly=0pt,bburx=530pt,bbury=530pt}
\caption{Comparison of the experimental data with events generated by the Monte-Carlo with the PWA model 
for $-\tprime < 0.08 \, \gev^2 $, $1.65 \, \gev \leq M(\fivpi) < 2.15 \, \gev$.
a) and g) $\pi^+\pi^-2\pi^0$, b) and h)$\pi^+2\pi^-\pi^0$, 
c) and i) $\pi^-\pi^0$.
d) Mass of $\tripi$. 
f) Azimuthal angle of the $\pi^-$-meson in the $\pi^-\pi^0$ centre of mass helicity
reference frame.
}
\label{pwa_mc_comp}
\end{figure}

	The real and imaginary parts of the density matrix are the parameters to be
defined by the extended maximum likelihood method \cite{Orear}. The fits with 
density matrix of different ranks showed that the matrix of the rank 3
is sufficient for the experimental data description. 
The quality of description is illustrated in
fig~\ref{pwa_mc_comp}, where some angular and invariant mass distributions for
events from one of the $M_{\ompipi}$ interval are plotted. 
The presented wave set was found to describe the data sufficiently well
up to the mass $M_{\ompipi} < 2.1 \, \gev$. The background model used
in the PWA turned out to be good and the wave FLAT was smooth and its intensity
was equal to the background in the $\omega$ peak region. The application of another
PWA method with the background subtraction by the variable $\lambda$, mentioned
above, proved the results of the PWA without subtraction. The results of the PWA of 
the $50 \, \mev$ and $100 \, \mev$ binned data were consistent.

\section{The results of the patial wave analysis} 

\begin{figure}[htpb]
\epsfig{file=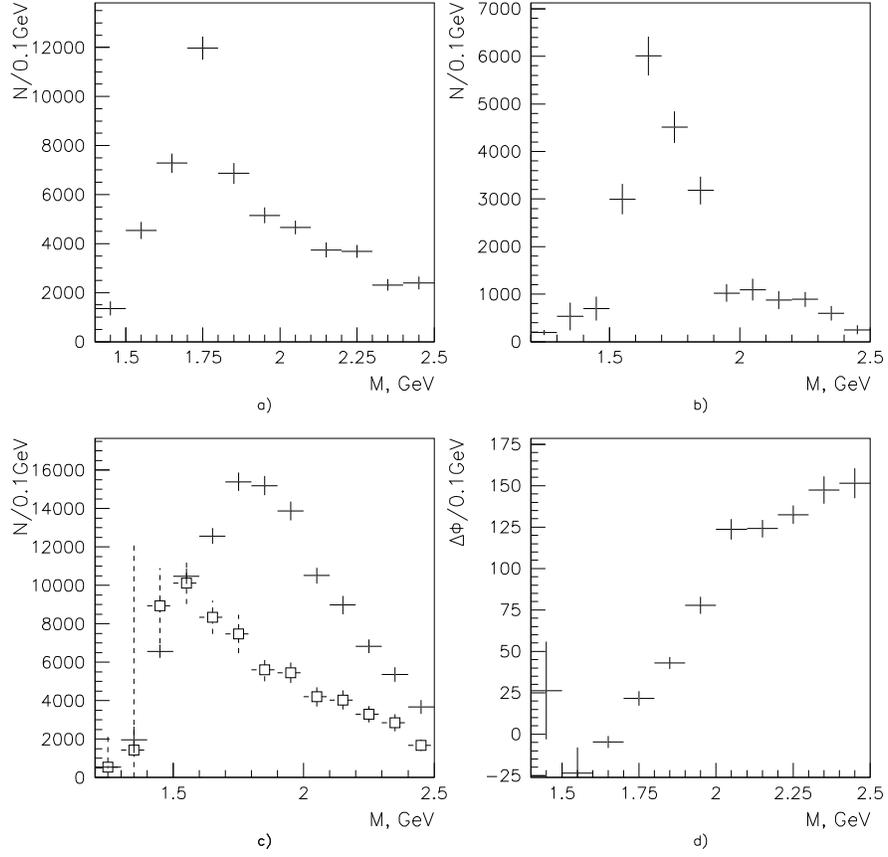,height=12cm,width=12cm,bbllx=0pt,bblly=0pt,bburx=530pt,bbury=530pt}
\caption{The PWA results for events with $-\tprime < 0.08 \, \gev^2 $. Wave 
intensities of:
a) $0^-0^+P1(\rho)$, 
b) $2^-0^+P2(\rho)$,
c) combined $1^+0^+$. Combined intensity of the $1^+0^+P(b_1) \mbox{ and } 1^+0^+S1(\rho)$
is drawn in squares. 
d) Phase of the $0^-0^+P1(\rho)$ wave with respect to the 
$1^+0^+P(b_1)$ wave. 
}
\label{waves_tot1}
\end{figure}

\begin{figure}[htpb]
\epsfig{file=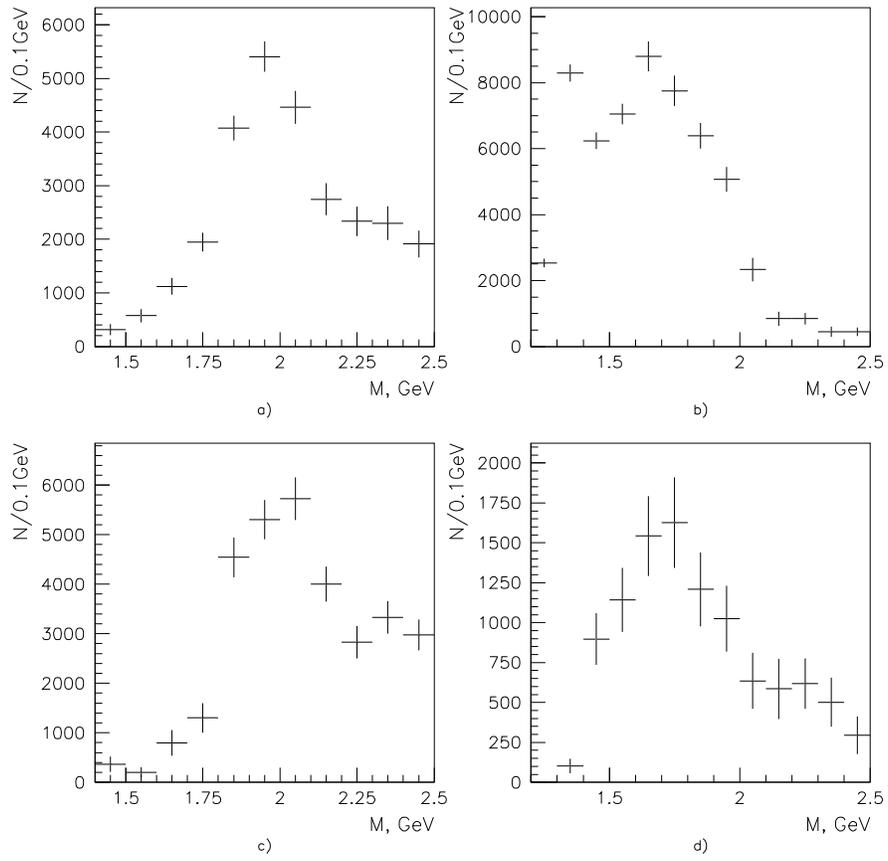,height=12cm,width=12cm,bbllx=0pt,bblly=0pt,bburx=530pt,bbury=530pt}
\caption{Wave intensities of: 
a) $3^+0^+S(\rho_3)$, 
b) $2^+1^+S2(\rho)$, 
c) $4^+1^+D2(\rho)$, 
d) $1^-1^+S(b_1)$.
}
\label{waves_tot2}
\end{figure}

	The results of the PWA are presented in fig.~\ref{waves_tot1} and ~\ref{waves_tot2}.

$\underline{\jpme=0^-0^+}$:
	A peak in the region of $1.74 \, \gev$ with the flat background 
dominates in the wave  $0^-0^+P(\rho)$ in the LT data sample 
(fig.~\ref{waves_tot1}a). 
The same structure is less visible in the $0^-0^+P(\rho)$ wave intensity
for events from the HT sample. The phase of these
waves with respect to the smooth waves $1^+0^+P(b_1)$, $1^+0^+S(\rho_1)$ exhibits 
resonant behaviour 
( fig.~\ref{waves_tot1}d). 
This fact makes it possible to assume the existence of a resonant state 
in the $0^-0^+$ wave. 

	The resonance parameters were determined by the fit of the wave intensity with
the incoherent sum of a relativistic Breit-Wigner function and a cubic 
polynomial.
The parameter estimates are robust to the fit with a coherent sum
and to the changes of the fit region. The mass was found to be
$M=1.737 \pm 0.005 \pm 0.015 \, \gev$ and the width  
$\Gamma=0.259 \pm 0.019 \pm 0.06 \, \gev$. The number of resonance decays
observed in the current experiment was estimated as
$N(\pi^-(1740) \rightarrow \omega\rho^-) = ( 230 \pm 12 \pm 82 ) \cdot 10^3 $.
The mass and width of this resonance differ from that for $\pi(1800)$.
The reason for this is unknown. It is possible to suggest the existence of 
two objects of different nature: a hybrid $\pi(1800)$ and a 
$3^1S_0 \mbox{ } q \bar q$ state decaying into $\omega\rho$.

$\underline{\jpme=1^+0^+}$:
	The intensity of the sum of all $1^+0^+$ waves has its maximum at 
$1.75 \, \gev$. Some structure can be seen at $ \sim 1.5 \, \gev$
(fig.~\ref{waves_tot1}c). This total
wave is a result of the combination of strongly interfering waves:
$1^+0^+P(b_1)$, $1^+0^+S1(\rho)$, $ 1^+0^+S(\rho_1)$, 
$1^+0^+D1(\rho)$ and $1^+0^+D2(\rho)$. 
The data do not allow one to state that the phase of some wave has a resonant
behaviour in the region of $ \sim 1.5 \, \gev$  or $ 1.75 \, \gev$.

$\underline{\jpme=2^-0^+}$:
A clear peak is observed
in the region of mass $1.67 \, \gev$ with width $ \sim 0.2 \, \gev$
(fig.~\ref{waves_tot1}b). 
The resonant phase behaviour of the $2^-0^+P1(\rho)$ and $ 2^-0^+P2(\rho)$
waves with respect to the $1^+0^+P(b_1)$ wave makes it possible to conclude the
observation of a resonance decaying into $\omega\rho$. This signal is
nearly absent in the $b_1\pi$ system. The resonance parameters of the
$2^-0^+P2(\rho)$ peak were estimated in the same way as for $\pi(1740)$:
 $M=1.687 \pm 0.009 \pm 0.015 \, \gev \mbox{ and } 
\Gamma=0.168 \pm 0.043 \pm 0.053 \, \gev$. We identify this 
phenomenon as the decay of $\pi_2(1670)$ into $\omega\rho$.

	The partial branching ratio was found by normalization to
the decay $\pi_2(1670) \rightarrow f_2(1270)\pi$ \cite{Kachaev},
observed in the current experiment decay: 

\begin{eqnarray}
Br(\pi_2(1670)^- \rightarrow \omrho S=2)
=0.019\pm0.004\pm0.010 \nonumber\\ 
Br(\pi_2(1670)^- \rightarrow \omrho S=1)
=0.009\pm0.002\pm0.003 \nonumber
\end{eqnarray}

	The error of the normalization to the $f_2\pi$ decay is sufficiently
large due to the Deck effect\cite{Deck}. The upper limits of the partial branching ratios of the decays $\pi_2(1670) \rightarrow \pi\rho_1, 
\rho_1 \rightarrow \omega\pi $ and 
$\pi_2(1670) \rightarrow b_1 \pi $ at the $2\sigma$ confidence
level were set:

\begin{eqnarray}
Br(\pi_2(1670) \rightarrow \rho_1\pi)<0.0036 , \nonumber \\
Br(\pi_2(1670) \rightarrow b_1 \pi)<0.0019 . \nonumber
\end{eqnarray}

	It is worth mentioning that a relatively large probability of the 
$\pi_2(1670) \rightarrow \omega \rho $ decay is in agreement 
with the model dependent calculations of \cite{Close}.


$\underline{\jpme=3^+0^+}$:
A peak at $M_{5\pi} \sim 2 \, \gev$ of the width  $ \sim 0.35 \, \gev $ is clearly
seen in the $3^+0^+S(\rho_3(1690))$ wave for events with low $\tprime$(LT)
(fig.~\ref{waves_tot2}a).
The resonant phase motion was not observed. Such wave behaviour can be
attributed to the Deck effect process, characterized by the near threshold bump in the
$S$-wave. A $3^+0^+$ state of the $\omega\rho$ system is $\sim 3$ times
smaller at the intensity maximum. 

$\underline{\jpme=1^+1^+}$:
This wave in the $\omrho$ is non-significant and does not have narrow structures of the width $\Gamma \sim 0.2 \, \gev$.

$\underline{\jpme=2^+1^+}$:
This wave dominates at high $\tprime$(HT sample). The intensive production and
decay of the $\at$ is the main process at low masses (fig.~\ref{waves_tot2}b). 
Due to the strong interefence of the $b_1\pi$ and $\omrho$ states in the 
region of their thresholds only the partial branching
fraction of the $\at \rightarrow \ompipi$ was determined relative to the 
well-known $\at$ decay modes into $\eta\pi$ and $\rho\pi$. This
probability was found to be $Br(\at \rightarrow \ompipi)=( 5 \pm 1 ) \% $.
There exists some arbitrariness in the question of what should be
called the $\at$ due to the complicated form of the $\ompipi$ spectrum in the $2^+1^+$ 
state. For possible misleading we define the $\at$ partial width
as that of a Breit-Wigner function with the S-wave $\ompipi$ background. This decay was previously observed in the low statistics experiments \cite{PDG}.

	The nature of the $1.7 \, \gev$ mass structure is unknown. The simple
suggestion is that this peak is due to the existence of a resonance with a
mass $M\approx1.6 \, \gev$ and a width $\Gamma \approx 0.3 \, \gev$. The radial
excitation of the $\at$ meson should exist in this region. Nevertheless, it is 
worth saying that even in the absence of such a resonance the mass specrtum
of the $2^+1^+ S$ $\ompipi$ wave should have a complicated two bump shape due to the opening of $\omega\rho$ chanel \cite{Migdal}. Sufficiently good description of this two bump spectrum was obtained in the model of one
resonance. These possibilities are different in the phase motion.
The phase motion was measured in this experiment with large uncertainty and 
did not allow us to make a definite conclusion.
The fraction of the $b_1\pi$ production in the same state in the band
$1.6 < M_{5\pi} < 2 \, \gev $ is negligible.

$\underline{\jpme=4^+1^+}$:
There is a peak in the mass spectrum of $\omrho$ system with the spin 2 and
orbital momentum 2 at $\approx 2 \, \gev$ (fig.~\ref{waves_tot2}c).
The phase shows resonant behaviour.
This signal can be identified as $\aq$. The $\aq$ parameters were estimated by
the mass spectrum fit with the incoherent sum of a $D$-wave relativistic Breit-Wigner functon
and a polynomial background. They were found to be:  	
 $M=1.944 \pm 0.008 \pm 0.050 \, \gev \mbox{ and } \Gamma=0.324 \pm 0.026 \pm 0.075 \, \gev$.
The $\tprime$ dependence of $\aq$ is identical to that of $\at$.

$\underline{\jpme=1^-1^+}$:
This exotic wave has quantum numbers which are forbidden for $q \bar q$ states.
The intensity of the $1^-1^+S(b_1)$ wave is very small for the LT sample events. 
That for high $\tprime$ shows a wide bump with the maximum at  
$M \sim 1.6 \div 1.7 \, \gev$. The signal in the intensity of the 
$\jpme=1^-1^+$ wave of the $\eta'\pi$-system with close parameters was
observed  earlier in \cite{Gouz}. All these facts indicate to the possibility
of the existence of a wide resonance ($\Gamma \sim 0.5 \, \gev$) with mass
$M \approx 1.6 \, \gev $.
The maximum intensity of this wave does not exceed $15 \%$ of the  
$2^+1^+S2(\rho)$ wave intensity (fig.~\ref{waves_tot2}d).
Similar waves in the $\omrho$ are also small.

\section{Conclusions} 

	The PWA of the $\ompipi$-system produced in the reaction  \reactf with beam momentum $P_{\pi} = 37 \, \gev $ was performed.

	A resonance structure with $M=1.737 \pm 0.005 \pm 0.015 \, \gev$ and 
$\Gamma=0.259 \pm 0.019 \pm 0.06 \, \gev$ was observed in the $\jpme=0^- 0^+ $ 
wave. These parameters differ from  that of $\pi(1800)$ seen in the $2\pi^-\pi^+$
system. This fact may point out to the existence of two different objects
with close masses.
 
	The $\at^- \rightarrow \ompipi $ decay in the wave $ \jpme=2^+1^+$ was 
found to have relative probability 
$ Br(a_2(1320)^- \rightarrow \omega \pi^- \pi^0) = ( 5 \pm 1 ) \% $.  

	A wide bump of unknown nature at $M \approx 1.7 \, \gev$ was observed
in this wave.

	The $\pi_2(1670) \rightarrow \omega \rho$ decay was observed in the
$\jpme=2^-0^+$ wave. The partial decay fraction was found to be:
\begin{eqnarray}
Br(\pi_2(1670)^- \rightarrow \omrho ) = 0.027\pm0.004\pm 0.01 \nonumber 
\end{eqnarray}

	The $ a_4(2040) \rightarrow \omega \rho $ decay was observed in
$\jpme=4^+1^+$ wave. We  measured the $ a_4(2040) $ resonance parameters :
  $M=1.944 \pm 0.008 \pm 0.050 
  \, \gev \mbox{ and  } \Gamma=0.324 \pm 0.026 \pm 0.075 \, \gev$. 

\bigskip

This research was supported in part by Russian Foundation for Basic Researches
under grant no.98-02-16392 and INTAS-RFBR grant no.95-0267.

\end{document}